\documentclass[aps,prb,twocolumn,amsmath,amssymb,amsfonts,showpacs]{revtex4-1}

\usepackage{mathptmx} 
\usepackage{dcolumn}                    
\usepackage{bm}                        
\usepackage{graphicx}
\usepackage{times}
\usepackage{epstopdf}
\usepackage{bbold}
\usepackage{xcolor}
\usepackage{dsfont}
\usepackage{amsthm} 
\usepackage[]{subfigure}
\usepackage{hypernat}
\usepackage{url}

\usepackage[hyperindex,breaklinks]{hyperref}
\hypersetup{
     colorlinks=true,        
     citecolor=blue,    
     filecolor=blue,      		
     urlcolor=blue,           	
    runcolor=cyan,
}

\begin{document}

\newcommand{\bfk}{\mathbf{k}}
\newcommand{\bfKpm}{\mathbf{K_\pm}}

\newcommand{\eq}{Eq.}
\newcommand{\eqs}{Eqs.}
\newcommand{\cf}{\textit{cf. }}
\newcommand{\ie}{\textit{i.e. }}
\newcommand{\eg}{\textit{e.g. }}
\newcommand{\etal}{\emph{et al.}}
\def\i{\mathrm{i}}

\title{Time-dependent topological systems: a study of the Bott index}

\author{Daniele Toniolo}
\email{daniele.toniolo@fau.de}

\affiliation{Department Mathematik, Friedrich-Alexander-Universit\"{a}t Erlangen-N\"{u}rnberg, 
Erlangen, Germany}

\date{\today}


\begin{abstract}
The Bott index is an index that discerns among pairs of unitary matrices that can or cannot be
approximated by a pair of commuting unitary matrices. It has been successfully employed to describe 
the approximate integer quantization of the transverse conductance of a system described by a 
short-range, bounded and spectrally gapped Hamiltonian on a finite two dimensional lattice on a torus  
and to describe the invariant of the Bernevig-Hughes-Zhang 
model even with disorder. This paper shows the constancy in time  of the Bott index and the Chern number related to the time-evolved Fermi projection of a
thermodynamically large system described by a short-range and time-dependent Hamiltonian that is initially gapped. The general situation of a ramp of a time-dependent perturbation is considered, a 
section is dedicated to time-periodic perturbations.  
\end{abstract}
\maketitle


\section{Introduction}
A recent focus of the research in condensed matter has been the description of the topological properties of systems that are subjected to perturbations in particular to time-driven ones. 
Early works on this subject are the references 
\cite{Lindner_Galitski:2011, Oka_Aoki:2009, Kitagawa_Demler:2011}. These consider how a time-periodic 
drive can induce a topological phase in the context of the Bernevig-Hughes-Zhang model and for a 
two 
dimensional system of fermions on a honeycomb lattice with spectral gap respectively. 
More recent 
studies \cite{Foster_Yuzbashyan:2013, D'Alessio_Rigol:2015, Caio_Cooper:2015, Oka_Mitra:2015, 
Caio_Cooper:2016, Kehrein:2016} have realized that the Chern number is invariant under the unitary 
time-evolution of the system, moreover in a two-dimensional setting when the Chern number of the initial ground state and that of the 
ground 
state of the instantaneous Hamiltonian are different then the Hall conductance is no more 
quantized. This has been shown both for the case of a quench from an initial trivial state to a 
topological one in the Haldane model \cite{Caio_Cooper:2015, Caio_Cooper:2016} and in the case of 
a system of spinless fermions of a honeycomb lattice described by an initial gapped 
Hamiltonian that is subjected to the linear ramp of a 
periodic 
external electromagnetic field \cite{D'Alessio_Rigol:2015, Oka_Mitra:2015}. The topology of 
periodically driven systems has been intensively studied, an incomplete list of works includes: the 
references that have introduced and rigorously discussed the $ W $ invariant in two dimensions
\cite{Rudner_Levin_2013, Graf_Tauber_2018} and in any dimension with the use of a K-theoretic 
construction \cite{ Sadel_Schulz-Baldes_2017}, the study of the chiral case 
\cite{Fruchart_2016} and the study of the time-reversal invariant case \cite{Carpentier_2015}, for 
the experimental side see \cite{Linking_2018} and references therein.
The invariance of the topological properties of the ground states, in general degenerate, of gapped 
Hamiltonians with respect to local unitarity transformations has been studied by Hastings and Wen 
\cite{Hastings_Wen_2005}. Related ideas recently brought to an explicit formulation of the 
adiabatic theorem in the many body context \cite{Bachmann_De_Roeck_2017, Bachmann_De_Roeck_2018}. 
A different approach 
to the dynamics of topological systems is the study of the winding of 
the Pancharatnam phase developed by the authors of reference \cite{Budich_Heyl:2016}.

The Bott index is an index of matrices that has been employed in the condensed matter realm by 
Loring and Hastings \cite{Loring_Hastings:2010}. The index has been introduced in ref. 
\cite{Loring:1988}, and its mathematical and physical foundations have 
been studied in \cite{Hastings_Loring:2010, Hastings_Loring:2011}, see also \cite{Loring_2014} and 
references therein. The Bott index discerns among couple of unitary or almost unitary matrices that 
can or cannot be
approximated by a commuting pair. In this paper we will be concerned with the index of a couple of 
unitary or quasi-unitary matrices, while the index has been used also for triples of Hermitian 
matrices 
and  for other classes in the case of systems with symmetries, time reversal or 
particle hole. Accordingly a  general classification of the topological phases of any 
of the Altland-Zirnbauer symmetry classes has been developed in \cite{Hastings_Loring:2010, 
Hastings_Loring:2011}. The Bott index of the projected position matrices on a torus, as defined by 
eq. \eqref{proj_positions} below, states whether those matrices can or cannot be approximated by a 
couple of commuting matrices. This encodes an information about the localization properties of the Fermi  
projection $ P $. The existence of exponentially localized Wannier functions 
imply the vanishing of the Bott index; a vanishing Bott index implies a small variance of the 
Wannier functions with respect to the system's size \cite{Hastings_Loring:2011}. Recently the 
relation among the spread and the localization of Wannier functions has been investigated in 
\cite{Monaco_Panati_2017}.
The Bott index is well suited for numerical simulations being designed for finite 
systems, it also handles  the effects of disorder.

The main subject of this work is to show that the Bott index of the time-evolved Fermi projection of a two-dimensional system described by 
a finite ranged, bounded and initially gapped Hamiltonian is constant in time in the thermodynamic limit when the Hamiltonian is subjected to a perturbation in general time dependent that preserves the locality of the Hamiltonian.  A possible change of the index in a certain time scale 
is only a finite size effect, this is the content of eq. \eqref{timechange}. The result is model-independent within the stated hypothesis on the Hamiltonian. An analogous numerical result for a specific model limited to a time-periodic periodic perturbation was provided in ref. \cite{Ge_Rigol:2017}. Another result of this work is to show that the Chern number of the time-evolved Fermi projection is constant in time under the same hypothesis for the Hamiltonian as in the Bott index case but in an infinite 2-dimensional systems. 
This also proves that the Bott index and the 2-dimensional Chern number are equivalent both in a time-independent setting    
\cite{Hastings_Loring:2011, Toniolo_Bott_eq:2017, Ge_Rigol:2017}  and in a time-dependent one.

A related result regarding the topological order of a set 
of degenerate ground states has been obtained in the reference \cite{Bravy_Hastings_2006}.  

The structure of the paper is as follows: in section \ref{Chern_unitary_evolution} the invariance 
of the 2-dim. Chern number along the time-evolution is proved. The Bott 
index 
is introduced in section \ref{Bott_def} and  the requirements on the physical setting for the index 
 to be well defined are  
described. The spectral flow that shows the mechanism of a possible 
variation of the Bott index is discussed in the unitary and in the general case. In section 
\ref{Bott_in_time} the growth 
in time of the norm of the commutator among the time-evolved Fermi projection and the 
position operator is studied and it is shown that for finite range Hamiltonians it cannot give rise to a 
change of the Bott index. Some technical details are in the appendix \ref{appendix_spectrum}.


\section{Invariance of the 2-dimensional Chern number of the time-evolved Fermi projection} \label{Chern_unitary_evolution}
Let us consider a two dimensional insulating system of non interacting particles with $ N $ internal degrees of freedom on an infinite lattice, that for convenience we take equal to $ \mathds{Z}^2 $, described by a single-particle gapped 
Hamiltonian $ H $ of finite range $ R $ acting on the Hilbert space $ l^2(\mathds{Z}^2) \otimes \mathds{C}^N  $.
\begin{equation} \nonumber
H = \sum_{l,k=1}^N \sum_{\|n-m\| \le R} \mathcal{H}_{n,m,l,k} |n,l \rangle \langle m,k| 
\end{equation}
The finite range condition reads: if $ \|n-m\|>R $ then $ \mathcal{H}_{n,m,l,k}=0 $.
$ \{|n \rangle, n=(n_x,n_y) \in \mathds{Z}^2\} $  is the usual basis of $ l^2(\mathds{Z}^2) $ such that $ |n \rangle $ equals $ 1 $ at the site $ n $ and zero elsewhere. The Fermi level $ \mu $ is supposed to lie in an energy gap of $ H $. 
The expression of the Chern number of the Fermi projection $ P \equiv \chi(H \le \mu ) $, here defined by the functional calculus with $ \chi $ the characteristic function, suitable for an evaluation of this 
topological invariant in real space has been employed for example as eq. (19) of \cite{Bellissard_1994} and  more recently in the  
appendix C of Kitaev's \cite{Kitaev_2006}. See also eq. (7) of \cite{Prodan_2009} and
\cite{Prodan_Hughes_Bernevig_2010, Prodan_review_2011}.
\begin{equation}
\label{Chernnumber}
\mathrm{Chern}(P)\equiv 4\pi\mathrm{Im}\mathrm{Tr_{u. a.}} Q 
\left[X,P\right]P \left[Y,P\right]
\end{equation}
Note that there is no uniformity in the choice of the sign of the Chern number in the literature. 
$ \mathrm{Tr_{u. a.}} $ is the trace per unit area:  $ \mathrm{Tr_{u. a.}} 
\equiv \lim_{A \rightarrow \infty} \frac{\mathrm{Tr \chi_A}}{A}$, $ \chi_A $ denotes multiplication by the characteristic function of the area $ A $, that is $ 1 $ inside $ A $, zero outside.  $ Q $ is the projection orthogonal to $ P
$, $ Q \equiv \mathds{1}- P $. The operators $ X $ and $ Y $ in eq. \eqref{Chernnumber} are the 
position operators of $ l^2(\mathds{Z}^2) $. For a discussion of the convergence of the trace that defines the Chern number see the so called Sobolev condition described for example in \cite{Bellissard_1994, Schulz-Baldes_BdG}, for a reformulation of the Chern number using switch functions instead of position operators and the inclusion of a wider class of Hamiltonian with exponentially decreasing amplitudes see for instance \cite{Graf_2005, Graf_Tauber_2018}.
Another perspective with equivalent results for the quantization of the Hall conductance is that of 
Avron \textit{et al.} \cite{Avron_Seiler_Simon_1994} that considered the Fredholm index of an operator associated to a couple 
of projections.

The constancy in time of the Chern number when the unitary time-evolution of the system is taken into account has been already shown analytically in the ref. \cite{D'Alessio_Rigol:2015} for a spatially periodic system 
following a time-evolution where a periodic perturbation is turned on and in ref. 
\cite{Caio_Cooper:2015} through a numerical evidence for the Haldane model following a quench. 
In both cases the Schroedinger picture for the unitary evolution was employed. This means that, in 
the present notation, the 
quantity 
\begin{align} \label{Chern_t}
& \mathrm{Chern}(P(t,t_0))= \\
\nonumber
& = 4\pi\mathrm{Im}\mathrm{Tr_{u. a.}} 
 Q(t,t_0) \left[X,P(t,t_0)\right]P(t,t_0) \left[Y,P(t,t_0)\right] 
\end{align}
 has been studied, where $ P(t,t_0) \equiv U(t,t_0)P(t_0)U^\dagger(t,t_0) $ and shown to be independent of $ t $ with a local Hamiltonian $ H(t)$. $ U(t,t_0) $ is 
the 
unitary operator of time-evolution of the system satisfying: 
\begin{equation} \label{Sch_eq}
i \partial_t U(t,t_0) = H(t) U(t,t_0), \hspace{5mm} U(t_0,t_0)=\mathds{1}
\end{equation}
 For  a time-independent 
system the 
time-evolution is given by $ U(t,t_0)= e^{-i H (t-t_0)} $, then the invariance of the Chern number is 
manifest, the relevant fact is the invariance in the general case 
of a system with a time dependent 
Hamiltonian $ H(t) $, see eq. \eqref{ramp}. The conditions required for the invariance of the Chern number 
under unitary evolution according to ref. \cite{D'Alessio_Rigol:2015} are the locality  of the instantaneous Hamiltonian $ H(t) $ in eq. \eqref{Sch_eq}  and certain regularity properties of the ground state projector over the Brillouin zone. In the present setting I consider the instantaneous Hamiltonian $ H(t) $ finite range and in general gapped only at the initial time $ t_0 $, this ensures the regularity of the projections $ P(t_0)$ and $ Q(t_0) $ as discussed for example in \cite{Prodan_2009}. I present an alternative proof of the constancy in time of the Chern number, $ \mathrm{Chern}(P(t,t_0)) = \mathrm{Chern}(P(t_0)) $. To this purpose the Chern number is expressed with the aid of switch functions $ \Lambda_x $ and $ \Lambda_y $ defined as follows: it exists a positive integer $ M $ such that with $ x>M $, $  \Lambda_x(x)=1 $ and  with $ x<-M $, $  \Lambda_x(x)=0 $ and  $  \Lambda_x $ varying continuously in between. $ \Lambda_y $ is similarly defined. Using the functional calculus the operators $ \Lambda_x(X) $ and $ \Lambda_y(Y) $ are defined, namely  $ \Lambda_x(X)|n_x,n_y \rangle = \Lambda_x(n_x)|n_x,n_y \rangle $. With abuse of notation I will write in the following $  \Lambda_x $ and  $ \Lambda_y $ for the corresponding operators. According, for example, to \cite{Bellissard_1994,  Avron_Seiler_Simon_1994} 
\begin{equation}
\label{ChernSwitch}
\mathrm{Chern}(P)= 4\pi \mathrm{Im} \mathrm{Tr} Q 
\left[\Lambda_x,P\right] P \left[\Lambda_y,P\right] 
\end{equation}
The $\mathrm{Tr}$ is over the Hilbert space $ l^2(\mathds{Z}^2) $. In what follows: $ t_0= 0 $, $ P \equiv P(t_0)$ and $ P(t) \equiv P(t,t_0) $. We want to show the invariance of the Chern when $ P $ is replaced with $ P(t) $. The Chern number of a projection is well defined when the trace in eq. \eqref{ChernSwitch} is finite, in this case the Chern number turns out to be an integer. Projectors that have well defined Chern number and that are homotopically equivalent have the same Chern number, for a proof of this statement in the context of Fredholm-index theory see for example \cite{Simon_Operator_Theory}. $ P $ and $ P(t) $ are homotopically equivalent therefore the task is to show that the trace in eq. \eqref{ChernSwitch} is finite when replacing  $ P $ with $ P(t) $.  In what follows it is convenient  to consider a more general class of Hamiltonians than the finite range ones, namely the class of Hamiltonians with off diagonal elements falling exponentially fast: $ |\langle n | H| m \rangle | \le M e^{-\nu \|n-m\| }$, this class of operators is called local. Let us consider:
\begin{equation} \nonumber
\mathrm{Chern}(P(t))= 4\pi\mathrm{Tr} Q(t) 
\left[\Lambda_x,P(t)\right] P(t) \left[\Lambda_y,P(t)\right] 
\end{equation}
Defining the Heisenberg picture $ X_{H}(t)\equiv U^\dagger(t) X U(t)$, with the aid of functional calculus we have:
\begin{equation} \nonumber
 \Lambda_x(X_{H}(t))=U^\dagger(t) \Lambda_x(X) U(t)
\end{equation}
Therefore, dropping the time index of $ U(t) $, we have:
\begin{equation}
\label{ChernSwitch_t}
\mathrm{Chern}(P(t))= 4\pi \mathrm{Im} \mathrm{Tr} Q 
\left[U^\dagger \Lambda_x U,P \right]P \left[U^\dagger \Lambda_yU,P\right] 
\end{equation}
The equation of motion for $ X_{H}(t)$ is:
\begin{equation} \nonumber
i \frac{d}{dt} X_{H}(t) =  \left[X,H(t)\right]_{H}.
\end{equation}
where the explicit time dependence of the Hamiltonian has been put in evidence. With $ X=X(t_0) $ we 
have:
\begin{equation} \nonumber
 i \left( X_{H}(t) -X \right) = \int_0^t ds  \left[X,H(s)\right]_{H}
\end{equation}
A simple manipulation of eq. \eqref{ChernSwitch_t} leads to:
\begin{align} \label{Chern_Switch_t_Lambda}
&\mathrm{Chern}(P(t))= \\ \nonumber
& =4\pi\mathrm{Im} \mathrm{Tr} Q \left[(\Lambda_x + U^\dagger[ \Lambda_x, U]),P\right] P \left[(\Lambda_y + U^\dagger[ \Lambda_y, U]),P\right]  \\ \nonumber
\end{align}
We replace in the equation above:
\begin{equation}
 U^\dagger[ \Lambda_x, U]=-i\int_{0}^t ds  U^\dagger(s)\left[\Lambda_x,H(s)\right]U(s)
\end{equation}
and similarly for the $y$-factor. We note that the operator   $ \left[\Lambda_x,H(s)\right] $ is confined around the $ y $ axis, this follows from the definition of $ \Lambda_x $ and from the locality of $ H(s) $ with $ s \in [0,t]$, this is discussed for example in the references \cite{Avron_Seiler_Simon_1994, Graf_Tauber_2018}. We will denote this behavior saying that  $ \left[\Lambda_x,H(s)\right] $  is $ x$-confined, namely that it exists a positive constant $ a $ such that the operator $  \left[\Lambda_x,H(s)\right] e^{a|x|} $ is bounded, see lemma 4.4 of ref. \cite{Graf_Tauber_2018}. In the case of a finite range $ H(s) $ the $ x$-confinement of $ \left[\Lambda_x,H(s)\right] $ is easily understood.  
  
To prove that the trace in eq. \eqref{Chern_Switch_t_Lambda} is finite and well defined, namely basis-independent, we need to show that the operator that we are tracing out is trace class, see \cite{Simon_Operator_Theory} for the definition, to do so we show that it splits up as a sum of trace class operators, denoted $ I $, $II$, $III$ and $IV$.
\begin{equation} \nonumber
 I\equiv Q[\Lambda_x,P]P[\Lambda_y,P] 
\end{equation}
is trace class by hypothesis, this follows from the fact that the Fermi projection $ P $ of a gapped and short range, or local,  Hamiltonian is local, namely it has off diagonal components falling off exponentially fast: $ |\langle n | P| m \rangle | \le K e^{-\mu \|n-m\| }$. 
\begin{equation} \label{Op}
II \equiv -i \int_0^t ds Q[\Lambda_x,P]P U^\dagger(s)\left[\Lambda_y,H(s)\right]U(s)
\end{equation}
is also trace class in fact $ [\Lambda_x,P] $ and $ [\Lambda_y,H(s)] $ are respectively $ x $ and $ y $-confined therefore their product is trace class, since the product of a trace class operator and a bounded operator is trace class it follows that $ P U^\dagger [\Lambda_y,H(s)] [\Lambda_x,P] $ is trace class. Using the cyclic property of the trace it follows that $ [\Lambda_x,P] P U^\dagger [\Lambda_y,H(s)] $ is trace class. Using again the fact that the trace class operators are an ideal of the bounded operators we get that \eqref{Op} is trace class.
\begin{align} \nonumber
& III \equiv -i \int_0^t ds' Q[\Lambda_y,P]P U^\dagger(s')\left[\Lambda_x,H(s')\right]U(s') \\ \nonumber
& IV \equiv \\ \nonumber
& \int_0^t ds ds'QU^\dagger(s)\left[\Lambda_x,H(s)\right]U(s)PU^\dagger(s')\left[\Lambda_y,H(s')\right]U(s')
\end{align}
Applying a similar reasoning as for $ II $ to $ III $ and $ IV $ we obtain that both are trace class. 

The above trace class discussion together with the homotopy-equivalence of $ P $ and $ P(t) $ shows that the value of the Chern number of $ P(t) $ is constant in time under the hypothesis that the Hamiltonian of the system $ H(s) $ is spatially local with $ s\in [0,t] $. This also means that the sum of the contributions to the trace of the operators $ II $, $ III $, and $ IV$ is zero.

We note that in the reasoning above we have exchanged the time integration and the trace, this can be justified simply. We suppose that the time-dependence of the Hamiltonian is at least strongly-continuous, this ensures the existence of a dynamics for the system, namely the propagator $ U(t,t_0) $ of eq. \eqref{Sch_eq} is well defined, see paragraph X.12 of \cite{Reed_Simon_2} for a discussion. We can approximate the time-integration by a finite sum plus a small remainder. We can safely exchange the trace and the finite sum, moreover since we conclude that the contribution of the finite sum is independent from $ t $ that means the contribution of the remainder is vanishing.





\section{Recollecting the definition of the Bott index and its relation with the 2-dimensional Chern number} \label{Bott_def}
In the references  
\cite{Hastings_Loring:2011, Toniolo_Bott_eq:2017, Ge_Rigol:2017} it has been shown that for a 
finite many-body system of non-interacting particles described by a short-ranged, bounded and gapped Hamiltonian 
living on a lattice on a two-torus
the invariant of matrices called Bott index coincides in the thermodynamic limit with the Chern 
number. In finite systems the correction is of order $ L^{-1} $, being $ L $ the linear size of the 
system. This in 
particular implies the quantization of the Hall conductance on a torus of finite size with an error 
of order $ L^{-1} $. This has been shown in the ref. \cite{Hastings_Loring:2011} exploiting 
the definition of the Hall conductance as the long time transverse current response of the system 
to an electric field adiabatically turned. The proof of ref. \cite{Toniolo_Bott_eq:2017} relies instead on a direct comparison of the invariants as they appear in eq. \eqref{Chernnumber} and eq. \eqref{Bott_PQ}.

Following ref. \cite{Hastings_Loring:2010} we consider a representation of the position operators $ 
X $ and $ Y $ such that the positions of all the particles of the system $ (x_i, y_i) $ are 
disposed on the diagonal of $ X $ and $ Y $ respectively: $ X_{i,j}=x_i \delta_{i,j} $. Being $ L $ 
the linear size of the system and assuming a lattice spacing equal to 1, $ X $ and $ Y $ are 
matrices of size of the order of $N L^2 \times N L^2 $. Denoting as in section \ref{Chern_unitary_evolution} the Fermi projection with $ P $ 
\begin{align}
\label{projectionP}
& P= W
\begin{pmatrix}
& 0 & 0 &\\
& 0 & \mathds{1}_n &\\
\end{pmatrix} 
W^\dagger 
\end{align}
with $ n = \dim P $ and $ W $ the unitary matrix of basis change from energy to position, so we also have: 
\begin{align}
\label{projectionQ}
 Q\equiv\mathds{1}-P = W
 \begin{pmatrix}
& \mathds{1}_m & 0 \\
& 0 & 0 \\
\end{pmatrix} 
W^\dagger
\end{align}
In a system with periodic boundary conditions we consider the unitary matrices $ 
e^{\left(i\frac{2\pi X}{L}\right)} $ and $ e^{\left(i\frac{2\pi Y}{L}\right)} $, then for 
the projected position operators on a torus $ P 
e^{\left(i\frac{2\pi X}{L}\right)} P$ and $ 
P e^{\left(i\frac{2\pi Y}{L}\right)} P$,  we have:
\begin{align} \label{proj_positions}
& P e^{\left(i\frac{2\pi X}{L}\right)} P = W
\begin{pmatrix}
& 0 & 0 &\\
& 0 & U_1 &\\
\end{pmatrix} 
W^\dagger \\
& P e^{\left(i\frac{2\pi Y}{L}\right)} P = W
\begin{pmatrix}
& 0 & 0 & \\
& 0 & U_2 & \\
\end{pmatrix} 
W^\dagger
\end{align}
where $ U_1 $ and $ U_2 $ are non singular. 
With $ R \ll L $ the range of the Hamiltonian, $ J $ a bound for the norm of $ H $ and $ \Delta E 
$ the spectral gap of the Hamiltonian it turns 
out \cite{Hastings_Loring:2010, Hastings_Loring:2011} that:
\begin{align}
\label{normcommpositionH}
 &\|[X,H]\| \le O(R J) \\
\label{normcommpositionP}
 &\|\left[e^{\left(i\frac{2\pi X}{L}\right)},P\right] \| \le O\left( \frac{R J}{L \Delta E}\right)
\end{align}
A couple of words on these relations: eq. 
\eqref{normcommpositionH} follows from the fact that $ X $ is a diagonal matrix then $ [X,H] $ 
looks like the off diagonal part of $ H $ with elements multiplied by factors of modulus at most 
equal to $ R $ because $ H $ connects lattice points that are at most $ R $ far apart. A similar bound occurs in the case of local Hamiltonian on $ l^2(\mathds{Z}^2)$ where $ R \propto \frac{1}{\nu}$.  Eq. 
\eqref{normcommpositionP} follows from \eqref{normcommpositionH} writing $ P $ as a contour 
integral of the resolvent $ R(z) \equiv (z-H)^{-1} $, see \cite{Hastings_Loring:2010} and 
\cite{Toniolo_Bott_eq:2017} for more 
details. This implies that the matrices $ U_1 $ and 
$ U_2 $ almost commute and are quasi unitarity:
\begin{align} 
\label{normcommU_1U_2}
&\|\left[U_1,U_2\right] \| \le O\left(\frac{RJ}{L \Delta E}\right)^2 \\
\label{quasiunitary}
& \|U_a U_a^\dagger - \mathds{1}_n \| \le O\left(\frac{RJ}{L \Delta E}\right)^2,\hspace{5mm} a 
\in \{1,2\}  
\end{align}
For a discussion of these results see \cite{Hastings_Loring:2011}.
The Bott index of $ U_1 $ and $ U_2 $ is defined as \cite{Loring_Hastings:2010}:
\begin{equation}
\label{Bott_definition}
 \mathrm{Bott}(U_1,U_2) \equiv \frac{1}{2\pi }\mathrm{Im}\mathrm{Tr}\log \left(U_1U_2U_1^\dagger 
U_2^\dagger \right)
\end{equation}
The branch cut of the $ \log $ is assumed on the real negative axis then the definition is well 
posed when $ U_1U_2U_1^\dagger U_2^\dagger $ has no real negative eigenvalue.  
 $ \mathrm{Bott}(U_1, U_2)=0 $ if and only if $ U_1 $ and $ U_2 $ are 
arbitrarily close 
to a couple of commuting quasi unitaries \cite{Loring_Hastings:2010}. This has been shown to be in 
relation with the existence of exponentially localized Wannier functions. More precisely the 
existence of exponentially localized Wannier functions implies the vanishing of the Bott index, 
while the vanishing of the Bott implies a spread of the Wannier functions, quantified by their 
variance, that is small compared with the linear size of the system,  
\cite{Hastings_Loring:2010, Hastings_Loring:2011}. 
An equivalent definition of the Bott index is given employing the matrices $ V_1 \equiv Q+ P 
e^{\left(i\frac{2\pi X}{L}\right)} P  $ and $ V_2 \equiv Q+ P 
e^{\left(i\frac{2\pi Y}{L}\right)} P $, then:
\begin{equation} \label{Bott_PQ}
 \mathrm{Bott}(U_1,U_2)=\frac{1}{2\pi }\mathrm{Im}\mathrm{Tr}\log \left(V_1V_2V_1^\dagger 
V_2^\dagger \right)
\end{equation}
The proof of the equivalence is immediate using the representation of the projections $ P $ and $ Q 
$, eqs. \eqref{projectionP}, \eqref{projectionQ}.

With the use of conditions \eqref{normcommU_1U_2} and \eqref{quasiunitary} it is possible to show 
that: 
\begin{equation}
\mathrm{Bott}(U_1,U_2)=\frac{1}{2\pi }\mathrm{Im}\mathrm{Tr}\left(P e^{i\theta_x} P e^{i\theta_y} 
P e^{-i\theta_x} P e^{-i\theta_y} P\right) + O(L^{-2})
\end{equation}
with $ \theta_x \equiv \frac{2\pi X}{L} $ and $ \theta_y \equiv \frac{2\pi Y}{L} $.
This expression is particularly well suited for numerical investigations.

\begin{figure}[ht]
   \centering
    \subfigure{%
    \includegraphics[width=0.48\linewidth]{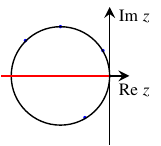} %
     }%
   \subfigure{%
    \includegraphics[width=0.48\linewidth]{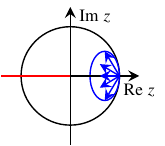}%
    }%
\caption{(Color on line) Left panel: with $ U_1 $ and $ U_2 $ unitary the spectrum of $ U_1U_2  
U_1^\dagger  U_2^\dagger -1 $ is a set of points 
on the black circle of radius equal to 1. $ \det (U_1U_2  
U_1^\dagger  U_2^\dagger) =1$.  The only 
point at a distance equal to 2 
from the origin is on the real negative axis. The thick red line indicates the branch cut of the 
log.   Right panel: the spectrum of  $ V_1 V_2 V_1^\dagger  V_2^\dagger  $ as given in eq. \eqref{Bott_PQ} 
is confined in a small region close to $(1,0)$. The unitary time-evolution of the system 
is such that $ P $ and $ Q $ are replaced by $ U(t,t_0)P(t_0)U^\dagger(t,t_0) $ and $ 
U(t,t_0)Q(t_0)U^\dagger(t,t_0) $ as a consequence the eigenvalues of $ V_1(t) V_2(t) V_1^\dagger(t)  V_2^\dagger(t)   $ can move in time but nevertheless they stay confined within the unit disc in an area close by $(1,0)$ schematically drawn as an ellipse. The thick red line indicates the branch cut of the log.} 
\label{non_unitary_spectrum} 
\end{figure}

Let us consider how the Bott index varies starting with the simpler case 
of unitary matrices namely when $ U_1 $ and $ U_2 $ in eq. \eqref{Bott_definition} are replaced by 
unitaries. In this case the Bott index is well defined when $ \|[U_1, U_2] \|<2 $, because with $ 
U_1 $ and $ U_2 $ unitary  $U_1U_2U_1^\dagger  
U_2^\dagger $ is unitary as well, that means 
its spectrum lies 
 on the unit circle of the complex plane. It is easy to see that $ \| \left[U_1 , U_2 \right] \| = 
\|U_1U_2  U_1^\dagger  U_2^\dagger -1 \| $ then recalling that the operator norm of a matrix $ A $ 
is $ \| A \| \equiv \max \{|\lambda|, \lambda \in \sigma(A) \} $ 
we see in the left panel of Fig. \ref{non_unitary_spectrum} that $ \|U_1U_2  U_1^\dagger  
U_2^\dagger -1 \| = 2 $ if 
and 
only if $ -1 $ belongs to the spectrum of $ U_1U_2  U_1^\dagger  U_2^\dagger $. Then when an 
eigenvalue crosses the real negative axis, that corresponds to cross the branch cut of the $ \log 
$,  its phases changes of $ 2\pi $ then the Bott index \eqref{Bott_definition} changes. In the ref. 
\cite{Loring_2014} it is discussed how a deformation (a homotopy) of a 
couple of unitary matrices that preserves their unitarity can lead to a change of their Bott index 
 only if at an intermediate 
point $ \| \left[U_1 , U_2 \right] \|=2 $. Namely: given a homotopy $ t \rightarrow (U_t, V_t) $ 
with $ (U_t, V_t) $ unitary matrices $ \forall t \in [0,1] $ then $ \mathrm{Bott}(U_0,V_0) \neq 
\mathrm{Bott}(U_1,V_1) $  only if it exists $ \bar{t}\in [0,1] $ such that $ 
\|[U_{\bar{t}},V_{\bar{t}} ]\| =2 $. See \cite{Loring_2014} and references therein for more 
theorems about it. In the case of the definition eq. \eqref{Bott_definition} where $ U_1 $ and $ 
U_2 
$ are not unitary we must consider where  the spectrum of  $ U_1U_2  U_1^\dagger  U_2^\dagger $ 
is located. It is possible to see, appendix \eqref{appendix_spectrum} for the details, that in for a time-independent system the spectrum is located close by the point $ (1,0) $ of the complex plane, this follows 
directly from equations \eqref{normcommU_1U_2} and \eqref{quasiunitary}. The time 
evolution generated by $ H(t) $ such that the operator $ U(t,t_0) $ commutes with $ P(t_0) $ leaves the spectrum in 
the same region, when  instead $ U(t,t_0) $ does not commute with $ P(t_0) $ , as explained in the next section, the time evolution causes the 
eigenvalues to move within the disc of unit radius, see the right panel of Fig. 
\ref{non_unitary_spectrum}.

\section{Invariance of the Bott Index of the time-evolved Fermi projection} \label{Bott_in_time}
The Chern number is invariant under unitary evolution of a generic local and time 
dependent Hamiltonian. We ask the same question about the Bott index. A way to realize the 
time driving of an initial Hamiltonian $ H_i $ towards a final Hamiltonian $ H_f(t) $ is the ramp 
of a perturbation $ V(t) $ in general time-dependent.
 \begin{equation} \label{ramp}
  H_f(t)=H_i + r(t)V(t), \hspace{5mm} r(t)=
   \begin{cases}
    0, \hspace{2mm} t<t_0 \\
    v(t), \hspace{2mm} t_0 \le t \le t_1 \\
    1, \hspace{2mm} t>t_1
  \end{cases}
 \end{equation}
with $ v(t) $ a monotonic regular function interpolating between zero and one. When the slope of $ 
v(t) $ increases significantly the driving becomes a so called quantum quench. In this case the 
operator of 
unitary evolution from the initial Hamiltonian $ H_i $ to the 
 final Hamiltonian $ H_f = 
 H_i + V $ is given in the case of $ V $ time-independent by 
 \begin{equation}
  U(t,t_0)=U(t,t_1)U(t_1,t_0)=e^{-i(t-t_1)H_f} U(t_1,t_0)
 \end{equation}
 With
 \begin{equation} \label{quench_time}
 t_1-t_0 \ll \left( \Delta_{\psi(t_0)} V \right)^{-1}
 \end{equation}
 the operator $ U(t_1,t_0) \simeq \mathds{1} $.
 In eq. \eqref{quench_time} 
 \begin{equation}
 \Delta_{\psi(t_0)} V \equiv \sqrt{\langle \psi(t_0) |V^2 
 |\psi(t_0)\rangle - \left( \langle 
 \psi(t_0) |V |\psi(t_0)\rangle \right)^2}
 \end{equation}
 is the variance of the perturbing potential over the 
 initial state $ |\psi(t_0)\rangle $.  This is 
 discussed for example in \cite{Galindo_Pascual_2}.

Let us study the time evolution of the Bott index, this can be done in the 
Schroedinger picture replacing $ P(t_0) $ with $ U(t,t_0)P(t_0)U^\dagger(t,t_0) $ and $ Q(t_0) $ 
with $ U(t,t_0)Q(t_0)U^\dagger(t,t_0) $. 
The invariance for a time-independent system is manifest, in fact  $ 
P(t_0) $ and $ U(t,t_0)=e^{-i(t-t_0)H} $ commute. I stress that this is different from considering 
the instantaneous Bott index that we would get replacing $ P $ with $ P(t)= \sum_i |\psi_i (t) 
\rangle \langle \psi_i (t) |$ being $ |\psi_i (t) \rangle $ the instantaneous eigenvector of the 
Hamiltonian $ H(t) |\psi_i (t) \rangle = E_i(t) |\psi_i (t) \rangle $, $  E_i(t) \le \mu $.   Is the 
invariance of the Bott index of the time-evolved Fermi projection also granted for a general time-dependent system with Hamiltonian \eqref{ramp}? 
A variation of the Bott index has been numerically shown in the ref. \cite{D'Alessio_Rigol:2015}, 
in the rest of this section I show that this can happen only as a finite size effect, the Bott 
index does not change due to the unitary evolution generated by a local Hamiltonian in the 
thermodynamic 
limit.

The analysis of the unitary case showed that the variation of the Bott index is due to the growth 
of $ \|[U_1,U_2]\| $. In our physical context the matrices $ U_1 $ and $ U_2 $ are not unitary 
and the increase of their commutator with time is due to the growth of $ \| [ 
e^{i\theta_x} , U(t,t_0)H(t_0)U^\dagger(t,t_0) ] \| = \| [ e^{i\theta_{x,H}(t)},H]\|$ with 
$\theta_{x,H}(t)\equiv U^\dagger(t,t_0)\theta_x U(t,t_0)$ the Heisenberg picture of $ \theta_x $. 
This determines in principle the growth of $ \| [ e^{i\theta_x} , U(t,t_0)PU^\dagger(t,t_0)] \| = 
\| [ 
e^{i\theta_{x,H}(t)},P]\|$ in time. Let us examine this explicitly. 

A time-independent, short-ranged, bounded and gapped Hamiltonian implies 
in a system large compared to the range $ R $ the relations \eqref{normcommpositionH}, 
\eqref{normcommpositionP}, 
\eqref{normcommU_1U_2} and \eqref{quasiunitary} above. In the general setting of a time-dependent 
Hamiltonian, that might be associated to the ramp of a time-dependent perturbation, the unitary 
operator of time evolution, also called the propagator, is: $ U(t,t_0)= T \exp \left(-i\int_{t_0}^t 
ds H(s) \right) $, $ T $ 
denotes the operator of time ordering. $ U(t,t_0) $ does not commute in general with the 
Hamiltonian $ H(t) $.  A possible way to examine $ \| [ U^\dagger(t,t_0) X U(t,t_0), H ] \| $ is to 
use the 
Lieb-Robinson bounds \cite{Lieb_Robinson:1972, Hastings_Les_Houches:2010} but these are better 
suited for operators that have  supports that do not overlap, therefore I 
employ a different strategy. 

The equation of motion for $ X_{H}(t)\equiv U^\dagger(t,t_0) X U(t,t_0)$ is:
$ i \frac{d}{dt} X_{H}(t) =  \left[X,H(t)\right]_{H} $.
The explicit time dependence of the Hamiltonian has been put in evidence. With $ X(t_0)=X $ we 
have:
\begin{equation} \nonumber
 i \left( X_{H}(t) -X \right) = \int_{t_0}^t ds  \left[X,H(s)\right]_{H}
\end{equation}
Being $ \| \left[X, U(t,t_0) \right] \| = \| U^\dagger(t,t_0) X U(t,t_0) - X \| $ it follows that:
\begin{equation} \nonumber
   \| \left[X, U(t,t_0) \right] \| \le |t-t_0| \sup_{s \in [t_0, t]} \| \left[X,H(s)\right] \|
\end{equation}
using eq. \eqref{normcommpositionH} and denoting $ R(t) $ the range of $ H(t) $ and $ J(t) $ its 
norm, it follows: 
\begin{equation} \label{commtime}
   \| \left[X, U(t,t_0) \right] \| \le |t-t_0| \sup_{s \in [t_0, t]}2 R(s)J(s)
\end{equation}

Our interest is in the growth with time of $ \| [ U^\dagger(t,t_0) X U(t,t_0), H(t_0) ] \| $. It 
follows from eq. \eqref{normcommpositionH} and eq. \eqref{commtime}, we drop the time-indexes of $ 
U(t,t_0) $ and set $ J \equiv J(t_0) $, that:
\begin{align}
 \| [ U^\dagger X U, H(t_0) ] \|& = \| [ U^\dagger \left[ X, U\right] + X, H(t_0) ] \|  \nonumber \\
&=\| [ U^\dagger \left[ X, U\right], H(t_0) ] + [X,H(t_0)] \| \nonumber \\
&\le \| [ U^\dagger \left[ X, U\right], H(t_0) ] \| + 2RJ \nonumber \\
&\le 2J\|[X,U] \| + 2RJ \nonumber \\
&\le 4 J|t-t_0| \sup_{s \in [t_0, t]} \left(R(s)J(s)\right) + 2RJ
\end{align}
In analogy to the static case, where eq. \eqref{normcommpositionH} implied eq. 
\eqref{normcommpositionP}, in the time-dependent case we have, being $ P \equiv P(t_0) $ the Fermi 
projection: 
\begin{align} \label{norm_pos_proj_time}
 \frac{ \| [ U^\dagger X U, P ] \|}{L} \le \frac{4  |t-t_0| \sup_{s 
\in [t_0, t]} \left(R(s)J(s)\right)}{L} + \frac{2RJ}{L \Delta E} 
\end{align} 
A necessary condition for the change of the Bott index is:
\begin{equation} \nonumber
 \frac{ \| [ U^\dagger X U, P ] \|}{L} \simeq 1
\end{equation}
This corresponds to a lower bound for the time interval $ \bar{t} -t_0 $ that would give rise to a 
change of the index such that:
\begin{equation} \label{timechange}
 \bar{t} -t_0 \simeq \frac{L - \frac{2 RJ}{\Delta E}}{4 \sup_{s \in [t_0, \bar{t}]} \left( R(s)J(s) \right)} 
\end{equation}
For a finite range Hamiltonian in the thermodynamic limit $ \frac{L}{R} \rightarrow \infty $ 
therefore if the instantaneous Hamiltonian $ H(t) $ has always a finite range $ R(t) $ the time 
scale for the variation of the index diverges. In a finite size setting the interplay of the ratios 
$ \frac{L}{R} \gg 1 $ and $ \frac{J}{\Delta E} \gg 1 $ might lead to an experimentally meaningful 
value of the time scale given eq. \eqref{timechange} that nevertheless corresponds to a lower bound 
 for the change of the Bott index, for this reason the present estimate does not dismiss the 
numerical results of the reference \cite{D'Alessio_Rigol:2015}.




\subsection{Periodically driven systems} \label{Floquet}
Let us consider the case in eq. \eqref{ramp} of the ramp of a time periodic perturbation $ 
V(t)=V(t+T) $.  This implies that the Hamiltonian $H(t)$ when $t\ge t_1 $ is time-periodic. In this 
case the 
estimate given by eq. \eqref{commtime} can be made sharper. In general the propagator $ U $  of a 
time-periodic Hamiltonian is not 
periodic, but when $ U $ has a spectral gap then there is a homotopy that maps $ U $ to 
a time-periodic propagator preserving the given gap as described in the reference 
\cite{Rudner_Levin_2013}. The Hamiltonian that generates this periodic propagator is called the 
relative Hamiltonian, its construction is described for example in 
\cite{Rudner_Levin_2013, Graf_Tauber_2018, Sadel_Schulz-Baldes_2017}. The propagator $ U $ and the 
homotopically equivalent time-periodic propagator share the same topological index $ W^\varepsilon 
$ of \cite{Rudner_Levin_2013}, see \cite{Graf_Tauber_2018} 
for a different naming, that 
characterizes each 
spectral gap of the propagator placed at $ e^{-i T \varepsilon} $ and therefore the Chern number of 
the spectral projection in between the various gaps. In fact denoting $ P^{\varepsilon,\varepsilon'} 
$ the spectral projection for the spectrum of $ U $ in between the gaps $ e^{-i T \varepsilon} $ and 
$ e^{-i T \varepsilon'} $ it holds: $ W^\varepsilon - W^{\varepsilon '} =  
\mathrm{Chern}(P^{\varepsilon',\varepsilon})  $. This is eq. 14 of \cite{Rudner_Levin_2013} or eq. 
3.22 of \cite{Graf_Tauber_2018}. 

The propagator $ U $ of a time periodic Hamiltonian of period $ T $ is such that $ \forall n \in 
\mathds{Z}$: $ U(t+nT,t_1+nT) = U(t,t_1)$. This property together with the supposed periodicity $ 
U(t_1 + T,t_1)=U(t_1,t_1)=\mathds{1}$ imply that $ U(t_1+nT,t_1)=\mathds{1}$. Therefore given $ t $ 
and $ t_1 $, since it exists a positive integer $ N $ such that $ t-(t_1 +NT) < T $, we have that: 
$U(t,t_1)=U(t,t_1+NT)U(t_1+NT,t_1)=U(t,t_1+NT)$.   
We suppose that according to eq. \eqref{ramp} the time needed to fully turn on the time-periodic 
perturbation $ V(t)=V(t+T)$ is equal to $ t_1-t_0 $ then the propagator of the Hamiltonian of eq. 
\eqref{ramp} for a suitable $ N $, denoting $ U_{\mathrm{per}} $ the periodic propagator, obeys the 
decomposition: 
\begin{align}
U(t,t_0)&=U_{\mathrm{per}}(t,t_1)U_{\mathrm{ramp}}(t_1,t_0) \nonumber \\
&=U_{\mathrm{per}}(t,t_1+NT)U_{\mathrm{ramp}}(t_1,t_0) \nonumber \\
&=U_{\mathrm{per}}(t-NT,t_1)U_{\mathrm{ramp}}(t_1,t_0) \nonumber 
\end{align}
In this way we conclude that the time interval $ |t-t_0| $ of eq. \eqref{timechange} is less
than $ t_1+T-t_0 $. Therefore for a periodic driving the change 
of the Bott index is not only forbidden at any given fixed $  |t-t_0| $ in the thermodynamic limit 
but also disfavored with respect to the general case for a finite size setting. 

I stress that the relative Hamiltonian of a space-local Hamiltonian under the hypothesis of 
existence of a spectral gap for the propagator is also space-local. This has been show in 
proposition 5.6 of reference \cite{Graf_Tauber_2018}, see also the specific notion of locality 
employed in that reference.

This discussion ignores possible heating effects that might takes place in the context of periodic 
driving, nevertheless recent works indicate the stability of such a phases over almost 
exponentially long times \cite{Russomanno_Silva_2012, Abanin_2015,Abanin_2017}.


\section{Concluding discussion}
The Bott index introduced in the physics' realm in the references \cite{Loring_Hastings:2010, 
Hastings_Loring:2010, Hastings_Loring:2011} has been investigated in a general \eqref{ramp} time-dependent setting 
and the constancy of the index  of the time-evolved Fermi projection has been established in the thermodynamic limit over. The time scale of a possible change of the index is identified in eq. \eqref{timechange}, this is a mere finite size effect, in particular it 
 looks disfavored in the time-periodic case.

A fundamental issue unexplored in this work is the 
meaning as a physical quantity of the Bott index for a general time-dependent Hamiltonian. In fact if for a static system it is equivalent to the Chern number so it measures the Hall conductance, what about instead 
a general time-dependent system? We should recall that the Hall conductance is not the mean value of 
an operator over a state but a transport coefficient computed with the aid of the Kubo formula. The 
Hall conductance has been found  to be not quantized after a quench according to the references 
\cite{Caio_Cooper:2016, Oka_Mitra:2015, Kehrein:2016}. The physical meaning of the quantized Bott 
index for a general time-dependent setting remains to be investigated.

I conclude with some final comments about the literature. 
In ref. 
\cite{Refael:2015} the Bott index has been claimed to be the right suited invariant to study finite 
systems that are disordered and periodically driven being the counterpart of the winding number 
invariant W introduce in the ref. \cite{Rudner_Levin_2013} for the study of clean and 
thermodynamically large periodically driven systems. W counts the number of edge states supported 
by a periodically driven two dimensional system. Let me comment briefly on the relation among 
the references 
\cite{D'Alessio_Rigol:2015} and \cite{Rudner_Levin_2013}: one of the most interesting results 
of \cite{D'Alessio_Rigol:2015} is to show that after ramping up a circularly polarized electric 
field on a graphene sheet with a staggering sublattice potential the initial ground state evolves 
keeping a vanishing Chern number despite the fact that the ground 
state 
of the final periodic Hamiltonian has a non trivial Chern number. This was shown in section 
\ref{Chern_unitary_evolution} using the invariance of the Chern number of homotopically equivalent projections and discussing the trace class properties of the operator on the RHS of eq. \eqref{Chern_Switch_t_Lambda}. 
The ref. \cite{Rudner_Levin_2013} on the other hand considers a system that is already in a 
periodically 
driven regime disregarding the effects of the ignition of the drive.
The issues related to the 
preparation of a periodically driven systems are also discussed \eg in the 
reference \cite{Abanin_2016}.  
Finally it needs to 
be mentioned that all the effects associated with phonons and their coupling with electrons have 
been neglected, for a study that takes these phenomena into account in the context of a quench of 
topological phases see \cite{Werner:2017}.

%

\section{Acknowledgements} It is a pleasure to thank Yang Ge and Marcos Rigol 
for exchange of correspondence, Hermann Schulz-Baldes, Yosi Avron, Jacob Shapiro and Florian Dorsch for discussions.


\appendix \section{} \label{appendix_spectrum}

We want to estimate the eigenvalues with maximum modulus (that is the norm) and the minimum modulus 
 of the matrix $ Q+P e^{i\theta_{x,H}} P e^{i\theta_{y,H}} P e^{-i\theta_{x,H}} P 
e^{-i\theta_{y,H}} P $ that is argument of the log that defines the Bott index 
\eqref{Bott_PQ}. The subscript $ H $ indicates the Heisenberg picture.
For simplicity we start considering $Q+Pe^{i\theta_{x,H}}P$. This matrix is not 
normal, but it admits a singular value decomposition: $Q+Pe^{i\theta_{x,H}}P=T^\dagger 
DS$, $ T $ and $ S $ are unitary, $ D $ is the diagonal matrix of eigenvalues. Then the modulus 
square of the eigenvalues of $Q+Pe^{i\theta_{x,H}}P$ are the eigenvalues of  $ G\equiv
Q+Pe^{-i\theta_{x,H}}Pe^{i\theta_{x,H}}P $. In fact:
\begin{align}
 G &=(Q+Pe^{i\theta_{x,H}}P)^\dagger (Q+Pe^{i\theta_{x,H}}P) \\
 &= S^\dagger D^\dagger T T^\dagger DS= 
S^\dagger D^\dagger  DS  
\end{align}
In the energy basis: 
\begin{align}
P=
\begin{pmatrix}
& 0 & 0 & \\
& 0 & \mathds{1}_n & \\
\end{pmatrix},\hspace{2mm}
Q=
\begin{pmatrix}
& \mathds{1}_m & 0 &\\
& 0 & 0  &\\
\end{pmatrix}
\end{align}
Then it is immediate to see that:
\begin{align} \label{double_matrix}
G=
\begin{pmatrix}
& \mathds{1}_m & 0 & \\
& 0 & W_4^\dagger W_4 &
\end{pmatrix}
\end{align}
With $ W_4 $ the lower diagonal block of the unitary matrix $ W^\dagger e^{i\theta_{x,H}} W$
\begin{align}
 W^\dagger e^{i\theta_{x,H}} W=
 \begin{pmatrix}
  & W_1 & W_2 & \\
  & W_3 & W_4 &
 \end{pmatrix}
\end{align}
From equation \eqref{double_matrix} we get:
\begin{align}
 \| G \| = \max \{1, \|W_4^\dagger W_4 \| \}=1
\end{align}
Note that $ W_4 $ satisfies, for example, $ W_2^\dagger W_2 + W_4^\dagger W_4 = \mathds{1} $, then 
being $ W_2^\dagger W_2 $ semipositive definite we have $ \| W_4 \| \le 1 $. $ W_2 $ can be 
vanishing only when $ W $ commutes with $ e^{-i\theta_{x,H}} $, and moreover when $ U(t,t_0) $ 
commutes with $ H(t_0) $ that is not the case we are interested in here. 
Then it follows that: 
\begin{equation}
 \|Q + Pe^{i\theta_{x,H}}P\| =1 
\end{equation}
With a similar argument we also obtain that: 
\begin{equation}
 \| Q+P e^{i\theta_{x,H}} P e^{i\theta_{y,H}} P e^{-i\theta_{x,H}} P e^{-i\theta_{y,H}} P\| =1
\end{equation}
Let us investigate the eigenvalue of minimum modulus of $ Q+P e^{i\theta_{x,H}} P e^{i\theta_{y,H}} 
P e^{-i\theta_{x,H}} P e^{-i\theta_{y,H}} P $. We again start considering the matrix 
$Q+Pe^{i\theta_{x,H}}P$ for simplicity. Its eigenvalue of minimum modulus is the square root of the 
smallest eigenvalue of  $ G $ that I indicate with $ 
\lambda_S $. It is easy to see using the positivity of $ G $ and $ \|G \|=1 $ that: 
\begin{align}
 1-\lambda_S &=\|\mathds{1}-G \|= \|\mathds{1} -Q- Pe^{-i\theta_{x,H}}Pe^{i\theta_{x,H}}P\| \\
 &=\|P(\mathds{1} - e^{-i\theta_{x,H}}Pe^{i\theta_{x,H}} )P\| \\
 &=\|Pe^{-i\theta_{x,H}}Qe^{i\theta_{x,H}}P\| \\
 \label{norm_smallest}
 &=\|[P,e^{-i\theta_{x,H}}]Q[e^{i\theta_{x,H}},P]\|
\end{align}
The eq. \eqref{norm_smallest} is a clear indication that when $ [e^{i\theta_{x,H}},P] $
is small then $ \lambda_S $ is close to 1. This follows directly from the  equation \eqref{normcommpositionP} that in turn follows from the 
hypothesis on the Hamiltonian's properties: short ranged, bounded, gapped. We have seen that a 
result of the time-evolution is to make the norm  $ \| [e^{i\theta_{x,H}},P] \|$ 
growing, this determines the decrease of $ \lambda_S $. We guess that when $ \| 
[e^{i\theta_{x,H}},P] 
\| \simeq 1$ then $ \lambda_S $ is small. We have seen that this is forbidden in the thermodynamic 
limit $ L \rightarrow \infty $. A similar analysis brings to the same conclusions for the 
eigenvalue of minimum modulus of $ Q+P e^{i\theta_{x,H}} P e^{i\theta_{y,H}} 
P e^{-i\theta_{x,H}} P e^{-i\theta_{y,H}} P $. As explained in the main text the eigenvalues of 
this operator are at the beginning all close to the point $ (1,0) $ of the complex plane. An 
equilibrium evolution would let them to stay in that region in such a way that the Bott index does 
not change. The time-evolution generated by a time-dependent Hamiltonian makes them to move within the disc of unit 
radius but nevertheless in the thermodynamic limit they do not cross the real negative axis in such a way the Bott index in invariant along the time-unitary evolution of the system.

\bibliography{biblio}

\begin{thebibliography}{47}%
\makeatletter
\providecommand \@ifxundefined [1]{%
 \@ifx{#1\undefined}
}%
\providecommand \@ifnum [1]{%
 \ifnum #1\expandafter \@firstoftwo
 \else \expandafter \@secondoftwo
 \fi
}%
\providecommand \@ifx [1]{%
 \ifx #1\expandafter \@firstoftwo
 \else \expandafter \@secondoftwo
 \fi
}%
\providecommand \natexlab [1]{#1}%
\providecommand \enquote  [1]{``#1''}%
\providecommand \bibnamefont  [1]{#1}%
\providecommand \bibfnamefont [1]{#1}%
\providecommand \citenamefont [1]{#1}%
\providecommand \href@noop [0]{\@secondoftwo}%
\providecommand \href [0]{\begingroup \@sanitize@url \@href}%
\providecommand \@href[1]{\@@startlink{#1}\@@href}%
\providecommand \@@href[1]{\endgroup#1\@@endlink}%
\providecommand \@sanitize@url [0]{\catcode `\\12\catcode `\$12\catcode
  `\&12\catcode `\#12\catcode `\^12\catcode `\_12\catcode `\%12\relax}%
\providecommand \@@startlink[1]{}%
\providecommand \@@endlink[0]{}%
\providecommand \url  [0]{\begingroup\@sanitize@url \@url }%
\providecommand \@url [1]{\endgroup\@href {#1}{\urlprefix }}%
\providecommand \urlprefix  [0]{URL }%
\providecommand \Eprint [0]{\href }%
\providecommand \doibase [0]{http://dx.doi.org/}%
\providecommand \selectlanguage [0]{\@gobble}%
\providecommand \bibinfo  [0]{\@secondoftwo}%
\providecommand \bibfield  [0]{\@secondoftwo}%
\providecommand \translation [1]{[#1]}%
\providecommand \BibitemOpen [0]{}%
\providecommand \bibitemStop [0]{}%
\providecommand \bibitemNoStop [0]{.\EOS\space}%
\providecommand \EOS [0]{\spacefactor3000\relax}%
\providecommand \BibitemShut  [1]{\csname bibitem#1\endcsname}%
\let\auto@bib@innerbib\@empty
\bibitem [{\citenamefont {Lindner}\ \emph {et~al.}(2011)\citenamefont
  {Lindner}, \citenamefont {Refael},\ and\ \citenamefont
  {Galitski}}]{Lindner_Galitski:2011}%
  \BibitemOpen
  \bibfield  {author} {\bibinfo {author} {\bibfnamefont {N.~H.}\ \bibnamefont
  {Lindner}}, \bibinfo {author} {\bibfnamefont {G.}~\bibnamefont {Refael}}, \
  and\ \bibinfo {author} {\bibfnamefont {V.}~\bibnamefont {Galitski}},\ }\href
  {\doibase 10.1038/nphys1926} {\bibfield  {journal} {\bibinfo  {journal}
  {Nature Physics}\ }\textbf {\bibinfo {volume} {7}},\ \bibinfo {pages} {490}
  (\bibinfo {year} {2011})}\BibitemShut {NoStop}%
\bibitem [{\citenamefont {Oka}\ and\ \citenamefont
  {Aoki}(2009)}]{Oka_Aoki:2009}%
  \BibitemOpen
  \bibfield  {author} {\bibinfo {author} {\bibfnamefont {T.}~\bibnamefont
  {Oka}}\ and\ \bibinfo {author} {\bibfnamefont {H.}~\bibnamefont {Aoki}},\
  }\href {\doibase 10.1103/PhysRevB.79.081406} {\bibfield  {journal} {\bibinfo
  {journal} {Phys. Rev. B}\ }\textbf {\bibinfo {volume} {79}},\ \bibinfo
  {pages} {081406} (\bibinfo {year} {2009})}\BibitemShut {NoStop}%
\bibitem [{\citenamefont {Kitagawa}\ \emph {et~al.}(2011)\citenamefont
  {Kitagawa}, \citenamefont {Oka}, \citenamefont {Brataas}, \citenamefont
  {Fu},\ and\ \citenamefont {Demler}}]{Kitagawa_Demler:2011}%
  \BibitemOpen
  \bibfield  {author} {\bibinfo {author} {\bibfnamefont {T.}~\bibnamefont
  {Kitagawa}}, \bibinfo {author} {\bibfnamefont {T.}~\bibnamefont {Oka}},
  \bibinfo {author} {\bibfnamefont {A.}~\bibnamefont {Brataas}}, \bibinfo
  {author} {\bibfnamefont {L.}~\bibnamefont {Fu}}, \ and\ \bibinfo {author}
  {\bibfnamefont {E.}~\bibnamefont {Demler}},\ }\href {\doibase
  10.1103/PhysRevB.84.235108} {\bibfield  {journal} {\bibinfo  {journal} {Phys.
  Rev. B}\ }\textbf {\bibinfo {volume} {84}},\ \bibinfo {pages} {235108}
  (\bibinfo {year} {2011})}\BibitemShut {NoStop}%
\bibitem [{\citenamefont {Foster}\ \emph {et~al.}(2013)\citenamefont {Foster},
  \citenamefont {Dzero}, \citenamefont {Gurarie},\ and\ \citenamefont
  {Yuzbashyan}}]{Foster_Yuzbashyan:2013}%
  \BibitemOpen
  \bibfield  {author} {\bibinfo {author} {\bibfnamefont {M.~S.}\ \bibnamefont
  {Foster}}, \bibinfo {author} {\bibfnamefont {M.}~\bibnamefont {Dzero}},
  \bibinfo {author} {\bibfnamefont {V.}~\bibnamefont {Gurarie}}, \ and\
  \bibinfo {author} {\bibfnamefont {E.~A.}\ \bibnamefont {Yuzbashyan}},\ }\href
  {\doibase 10.1103/PhysRevB.88.104511} {\bibfield  {journal} {\bibinfo
  {journal} {Phys. Rev. B}\ }\textbf {\bibinfo {volume} {88}},\ \bibinfo
  {pages} {104511} (\bibinfo {year} {2013})}\BibitemShut {NoStop}%
\bibitem [{\citenamefont {D'Alessio}\ and\ \citenamefont
  {Rigol}()}]{D'Alessio_Rigol:2015}%
  \BibitemOpen
  \bibfield  {author} {\bibinfo {author} {\bibfnamefont {L.}~\bibnamefont
  {D'Alessio}}\ and\ \bibinfo {author} {\bibfnamefont {M.}~\bibnamefont
  {Rigol}},\ }\href {\doibase 10.1038/ncomms9336} {\bibfield  {journal}
  {\bibinfo  {journal} {Nature Communications}\ }\textbf {\bibinfo {volume}
  {6}},\ \bibinfo {pages} {8336}}\BibitemShut {NoStop}%
\bibitem [{\citenamefont {Caio}\ \emph {et~al.}(2015)\citenamefont {Caio},
  \citenamefont {Cooper},\ and\ \citenamefont {Bhaseen}}]{Caio_Cooper:2015}%
  \BibitemOpen
  \bibfield  {author} {\bibinfo {author} {\bibfnamefont {M.~D.}\ \bibnamefont
  {Caio}}, \bibinfo {author} {\bibfnamefont {N.~R.}\ \bibnamefont {Cooper}}, \
  and\ \bibinfo {author} {\bibfnamefont {M.~J.}\ \bibnamefont {Bhaseen}},\
  }\href {\doibase 10.1103/PhysRevLett.115.236403} {\bibfield  {journal}
  {\bibinfo  {journal} {Phys. Rev. Lett.}\ }\textbf {\bibinfo {volume} {115}},\
  \bibinfo {pages} {236403} (\bibinfo {year} {2015})}\BibitemShut {NoStop}%
\bibitem [{\citenamefont {Dehghani}\ \emph {et~al.}(2015)\citenamefont
  {Dehghani}, \citenamefont {Oka},\ and\ \citenamefont
  {Mitra}}]{Oka_Mitra:2015}%
  \BibitemOpen
  \bibfield  {author} {\bibinfo {author} {\bibfnamefont {H.}~\bibnamefont
  {Dehghani}}, \bibinfo {author} {\bibfnamefont {T.}~\bibnamefont {Oka}}, \
  and\ \bibinfo {author} {\bibfnamefont {A.}~\bibnamefont {Mitra}},\ }\href
  {\doibase 10.1103/PhysRevB.91.155422} {\bibfield  {journal} {\bibinfo
  {journal} {Phys. Rev. B}\ }\textbf {\bibinfo {volume} {91}},\ \bibinfo
  {pages} {155422} (\bibinfo {year} {2015})}\BibitemShut {NoStop}%
\bibitem [{\citenamefont {Caio}\ \emph {et~al.}(2016)\citenamefont {Caio},
  \citenamefont {Cooper},\ and\ \citenamefont {Bhaseen}}]{Caio_Cooper:2016}%
  \BibitemOpen
  \bibfield  {author} {\bibinfo {author} {\bibfnamefont {M.~D.}\ \bibnamefont
  {Caio}}, \bibinfo {author} {\bibfnamefont {N.~R.}\ \bibnamefont {Cooper}}, \
  and\ \bibinfo {author} {\bibfnamefont {M.~J.}\ \bibnamefont {Bhaseen}},\
  }\href {\doibase 10.1103/PhysRevB.94.155104} {\bibfield  {journal} {\bibinfo
  {journal} {Phys. Rev. B}\ }\textbf {\bibinfo {volume} {94}},\ \bibinfo
  {pages} {155104} (\bibinfo {year} {2016})}\BibitemShut {NoStop}%
\bibitem [{\citenamefont {Wang}\ \emph {et~al.}(2016)\citenamefont {Wang},
  \citenamefont {Schmitt},\ and\ \citenamefont {Kehrein}}]{Kehrein:2016}%
  \BibitemOpen
  \bibfield  {author} {\bibinfo {author} {\bibfnamefont {P.}~\bibnamefont
  {Wang}}, \bibinfo {author} {\bibfnamefont {M.}~\bibnamefont {Schmitt}}, \
  and\ \bibinfo {author} {\bibfnamefont {S.}~\bibnamefont {Kehrein}},\ }\href
  {\doibase 10.1103/PhysRevB.93.085134} {\bibfield  {journal} {\bibinfo
  {journal} {Phys. Rev. B}\ }\textbf {\bibinfo {volume} {93}},\ \bibinfo
  {pages} {085134} (\bibinfo {year} {2016})}\BibitemShut {NoStop}%
\bibitem [{\citenamefont {Rudner}\ \emph {et~al.}(2013)\citenamefont {Rudner},
  \citenamefont {Lindner}, \citenamefont {Berg},\ and\ \citenamefont
  {Levin}}]{Rudner_Levin_2013}%
  \BibitemOpen
  \bibfield  {author} {\bibinfo {author} {\bibfnamefont {M.~S.}\ \bibnamefont
  {Rudner}}, \bibinfo {author} {\bibfnamefont {N.~H.}\ \bibnamefont {Lindner}},
  \bibinfo {author} {\bibfnamefont {E.}~\bibnamefont {Berg}}, \ and\ \bibinfo
  {author} {\bibfnamefont {M.}~\bibnamefont {Levin}},\ }\href {\doibase
  10.1103/PhysRevX.3.031005} {\bibfield  {journal} {\bibinfo  {journal} {Phys.
  Rev. X}\ }\textbf {\bibinfo {volume} {3}},\ \bibinfo {pages} {031005}
  (\bibinfo {year} {2013})}\BibitemShut {NoStop}%
\bibitem [{\citenamefont {Graf}\ and\ \citenamefont
  {Tauber}(2018)}]{Graf_Tauber_2018}%
  \BibitemOpen
  \bibfield  {author} {\bibinfo {author} {\bibfnamefont {G.~M.}\ \bibnamefont
  {Graf}}\ and\ \bibinfo {author} {\bibfnamefont {C.}~\bibnamefont {Tauber}},\
  }\href {\doibase 10.1007/s00023-018-0657-7} {\bibfield  {journal} {\bibinfo
  {journal} {Annales Henri Poincar{\'e}}\ }\textbf {\bibinfo {volume} {19}},\
  \bibinfo {pages} {709} (\bibinfo {year} {2018})}\BibitemShut {NoStop}%
\bibitem [{\citenamefont {Sadel}\ and\ \citenamefont
  {Schulz-Baldes}(2017)}]{Sadel_Schulz-Baldes_2017}%
  \BibitemOpen
  \bibfield  {author} {\bibinfo {author} {\bibfnamefont {C.}~\bibnamefont
  {Sadel}}\ and\ \bibinfo {author} {\bibfnamefont {H.}~\bibnamefont
  {Schulz-Baldes}},\ }\href {\doibase 10.1007/s11040-017-9253-1} {\bibfield
  {journal} {\bibinfo  {journal} {Mathematical Physics, Analysis and Geometry}\
  }\textbf {\bibinfo {volume} {20}},\ \bibinfo {pages} {22} (\bibinfo {year}
  {2017})}\BibitemShut {NoStop}%
\bibitem [{\citenamefont {Fruchart}(2016)}]{Fruchart_2016}%
  \BibitemOpen
  \bibfield  {author} {\bibinfo {author} {\bibfnamefont {M.}~\bibnamefont
  {Fruchart}},\ }\href {\doibase 10.1103/PhysRevB.93.115429} {\bibfield
  {journal} {\bibinfo  {journal} {Phys. Rev. B}\ }\textbf {\bibinfo {volume}
  {93}},\ \bibinfo {pages} {115429} (\bibinfo {year} {2016})}\BibitemShut
  {NoStop}%
\bibitem [{\citenamefont {Carpentier}\ \emph {et~al.}(2015)\citenamefont
  {Carpentier}, \citenamefont {Delplace}, \citenamefont {Fruchart},\ and\
  \citenamefont {Gawedzki}}]{Carpentier_2015}%
  \BibitemOpen
  \bibfield  {author} {\bibinfo {author} {\bibfnamefont {D.}~\bibnamefont
  {Carpentier}}, \bibinfo {author} {\bibfnamefont {P.}~\bibnamefont
  {Delplace}}, \bibinfo {author} {\bibfnamefont {M.}~\bibnamefont {Fruchart}},
  \ and\ \bibinfo {author} {\bibfnamefont {K.}~\bibnamefont {Gawedzki}},\
  }\href {\doibase 10.1103/PhysRevLett.114.106806} {\bibfield  {journal}
  {\bibinfo  {journal} {Phys. Rev. Lett.}\ }\textbf {\bibinfo {volume} {114}},\
  \bibinfo {pages} {106806} (\bibinfo {year} {2015})}\BibitemShut {NoStop}%
\bibitem [{\citenamefont {Tarnowski}()}]{Linking_2018}%
  \BibitemOpen
  \bibfield  {author} {\bibinfo {author} {\bibfnamefont {M.}~\bibnamefont
  {Tarnowski}, \bibfnamefont {{\it et al.}}},\ }\href
  {https://arxiv.org/abs/1709.01046} {\ }\Eprint
  {http://arxiv.org/abs/1709.01046} {arXiv:1709.01046 [cond-mat.quant-gas]}
  \BibitemShut {NoStop}%
\bibitem [{\citenamefont {Hastings}\ and\ \citenamefont
  {Wen}(2005)}]{Hastings_Wen_2005}%
  \BibitemOpen
  \bibfield  {author} {\bibinfo {author} {\bibfnamefont {M.~B.}\ \bibnamefont
  {Hastings}}\ and\ \bibinfo {author} {\bibfnamefont {X.-G.}\ \bibnamefont
  {Wen}},\ }\href {\doibase 10.1103/PhysRevB.72.045141} {\bibfield  {journal}
  {\bibinfo  {journal} {Phys. Rev. B}\ }\textbf {\bibinfo {volume} {72}},\
  \bibinfo {pages} {045141} (\bibinfo {year} {2005})}\BibitemShut {NoStop}%
\bibitem [{\citenamefont {Bachmann}\ \emph {et~al.}(2017)\citenamefont
  {Bachmann}, \citenamefont {De~Roeck},\ and\ \citenamefont
  {Fraas}}]{Bachmann_De_Roeck_2017}%
  \BibitemOpen
  \bibfield  {author} {\bibinfo {author} {\bibfnamefont {S.}~\bibnamefont
  {Bachmann}}, \bibinfo {author} {\bibfnamefont {W.}~\bibnamefont {De~Roeck}},
  \ and\ \bibinfo {author} {\bibfnamefont {M.}~\bibnamefont {Fraas}},\ }\href
  {\doibase 10.1103/PhysRevLett.119.060201} {\bibfield  {journal} {\bibinfo
  {journal} {Phys. Rev. Lett.}\ }\textbf {\bibinfo {volume} {119}},\ \bibinfo
  {pages} {060201} (\bibinfo {year} {2017})}\BibitemShut {NoStop}%
\bibitem [{\citenamefont {Bachmann}\ \emph {et~al.}(2018)\citenamefont
  {Bachmann}, \citenamefont {De~Roeck},\ and\ \citenamefont
  {Fraas}}]{Bachmann_De_Roeck_2018}%
  \BibitemOpen
  \bibfield  {author} {\bibinfo {author} {\bibfnamefont {S.}~\bibnamefont
  {Bachmann}}, \bibinfo {author} {\bibfnamefont {W.}~\bibnamefont {De~Roeck}},
  \ and\ \bibinfo {author} {\bibfnamefont {M.}~\bibnamefont {Fraas}},\ }\href
  {\doibase 10.1007/s00220-018-3117-9} {\bibfield  {journal} {\bibinfo
  {journal} {Communications in Mathematical Physics}\ }\textbf {\bibinfo
  {volume} {361}},\ \bibinfo {pages} {997} (\bibinfo {year}
  {2018})}\BibitemShut {NoStop}%
\bibitem [{\citenamefont {Budich}\ and\ \citenamefont
  {Heyl}(2016)}]{Budich_Heyl:2016}%
  \BibitemOpen
  \bibfield  {author} {\bibinfo {author} {\bibfnamefont {J.~C.}\ \bibnamefont
  {Budich}}\ and\ \bibinfo {author} {\bibfnamefont {M.}~\bibnamefont {Heyl}},\
  }\href {\doibase 10.1103/PhysRevB.93.085416} {\bibfield  {journal} {\bibinfo
  {journal} {Phys. Rev. B}\ }\textbf {\bibinfo {volume} {93}},\ \bibinfo
  {pages} {085416} (\bibinfo {year} {2016})}\BibitemShut {NoStop}%
\bibitem [{\citenamefont {Loring}\ and\ \citenamefont
  {Hastings}(2010)}]{Loring_Hastings:2010}%
  \BibitemOpen
  \bibfield  {author} {\bibinfo {author} {\bibfnamefont {T.~A.}\ \bibnamefont
  {Loring}}\ and\ \bibinfo {author} {\bibfnamefont {M.~B.}\ \bibnamefont
  {Hastings}},\ }\href {http://stacks.iop.org/0295-5075/92/i=6/a=67004}
  {\bibfield  {journal} {\bibinfo  {journal} {EPL (Europhysics Letters)}\
  }\textbf {\bibinfo {volume} {92}},\ \bibinfo {pages} {67004} (\bibinfo {year}
  {2010})}\BibitemShut {NoStop}%
\bibitem [{\citenamefont {Loring}(1988)}]{Loring:1988}%
  \BibitemOpen
  \bibfield  {author} {\bibinfo {author} {\bibfnamefont {T.~A.}\ \bibnamefont
  {Loring}},\ }\href {\doibase 10.4153/CJM-1988-008-9} {\bibfield  {journal}
  {\bibinfo  {journal} {Canad. J. Math.}\ }\textbf {\bibinfo {volume} {40}},\
  \bibinfo {pages} {197} (\bibinfo {year} {1988})}\BibitemShut {NoStop}%
\bibitem [{\citenamefont {Hastings}\ and\ \citenamefont
  {Loring}(2010)}]{Hastings_Loring:2010}%
  \BibitemOpen
  \bibfield  {author} {\bibinfo {author} {\bibfnamefont {M.~B.}\ \bibnamefont
  {Hastings}}\ and\ \bibinfo {author} {\bibfnamefont {T.~A.}\ \bibnamefont
  {Loring}},\ }\href {\doibase 10.1063/1.3274817} {\bibfield  {journal}
  {\bibinfo  {journal} {Journal of Mathematical Physics}\ }\textbf {\bibinfo
  {volume} {51}},\ \bibinfo {pages} {015214} (\bibinfo {year}
  {2010})}\BibitemShut {NoStop}%
\bibitem [{\citenamefont {Hastings}\ and\ \citenamefont
  {Loring}(2011)}]{Hastings_Loring:2011}%
  \BibitemOpen
  \bibfield  {author} {\bibinfo {author} {\bibfnamefont {M.~B.}\ \bibnamefont
  {Hastings}}\ and\ \bibinfo {author} {\bibfnamefont {T.~A.}\ \bibnamefont
  {Loring}},\ }\href {\doibase http://dx.doi.org/10.1016/j.aop.2010.12.013}
  {\bibfield  {journal} {\bibinfo  {journal} {Annals of Physics}\ }\textbf
  {\bibinfo {volume} {326}},\ \bibinfo {pages} {1699 } (\bibinfo {year}
  {2011})},\ \bibinfo {note} {july 2011 Special Issue}\BibitemShut {NoStop}%
\bibitem [{\citenamefont {Loring}()}]{Loring_2014}%
  \BibitemOpen
  \bibfield  {author} {\bibinfo {author} {\bibfnamefont {T.~A.}\ \bibnamefont
  {Loring}},\ }\href {\doibase 10.3842/SIGMA.2014.077} {\bibfield  {journal}
  {\bibinfo  {journal} {Sigma}\ }\textbf {\bibinfo {volume} {10}},\ \bibinfo
  {pages} {077}}\BibitemShut {NoStop}%
\bibitem [{\citenamefont {Monaco}\ \emph {et~al.}()\citenamefont {Monaco},
  \citenamefont {Panati}, \citenamefont {Pisante},\ and\ \citenamefont
  {Teufel}}]{Monaco_Panati_2017}%
  \BibitemOpen
  \bibfield  {author} {\bibinfo {author} {\bibfnamefont {D.}~\bibnamefont
  {Monaco}}, \bibinfo {author} {\bibfnamefont {G.}~\bibnamefont {Panati}},
  \bibinfo {author} {\bibfnamefont {A.}~\bibnamefont {Pisante}}, \ and\
  \bibinfo {author} {\bibfnamefont {S.}~\bibnamefont {Teufel}},\ }\href
  {https://arxiv.org/abs/1612.09557v1} {\ }\Eprint
  {http://arxiv.org/abs/1612.09557} {arXiv:1612.09557 [cond-mat]} \BibitemShut
  {NoStop}%
\bibitem [{\citenamefont {Ge}\ and\ \citenamefont
  {Rigol}(2017)}]{Ge_Rigol:2017}%
  \BibitemOpen
  \bibfield  {author} {\bibinfo {author} {\bibfnamefont {Y.}~\bibnamefont
  {Ge}}\ and\ \bibinfo {author} {\bibfnamefont {M.}~\bibnamefont {Rigol}},\
  }\href {\doibase 10.1103/PhysRevA.96.023610} {\bibfield  {journal} {\bibinfo
  {journal} {Phys. Rev. A}\ }\textbf {\bibinfo {volume} {96}},\ \bibinfo
  {pages} {023610} (\bibinfo {year} {2017})}\BibitemShut {NoStop}%
\bibitem [{\citenamefont {Toniolo}()}]{Toniolo_Bott_eq:2017}%
  \BibitemOpen
  \bibfield  {author} {\bibinfo {author} {\bibfnamefont {D.}~\bibnamefont
  {Toniolo}},\ }\href {https://arxiv.org/abs/1708.05912} {\ }\Eprint
  {http://arxiv.org/abs/1708.05912} {arXiv:1708.05912 [cond-mat]} \BibitemShut
  {NoStop}%
\bibitem [{\citenamefont {Bravyi}\ \emph {et~al.}(2006)\citenamefont {Bravyi},
  \citenamefont {Hastings},\ and\ \citenamefont
  {Verstraete}}]{Bravy_Hastings_2006}%
  \BibitemOpen
  \bibfield  {author} {\bibinfo {author} {\bibfnamefont {S.}~\bibnamefont
  {Bravyi}}, \bibinfo {author} {\bibfnamefont {M.~B.}\ \bibnamefont
  {Hastings}}, \ and\ \bibinfo {author} {\bibfnamefont {F.}~\bibnamefont
  {Verstraete}},\ }\href {\doibase 10.1103/PhysRevLett.97.050401} {\bibfield
  {journal} {\bibinfo  {journal} {Phys. Rev. Lett.}\ }\textbf {\bibinfo
  {volume} {97}},\ \bibinfo {pages} {050401} (\bibinfo {year}
  {2006})}\BibitemShut {NoStop}%
\bibitem [{\citenamefont {Bellissard}\ \emph {et~al.}(1994)\citenamefont
  {Bellissard}, \citenamefont {van Elst},\ and\ \citenamefont
  {Schulz-Baldes}}]{Bellissard_1994}%
  \BibitemOpen
  \bibfield  {author} {\bibinfo {author} {\bibfnamefont {J.}~\bibnamefont
  {Bellissard}}, \bibinfo {author} {\bibfnamefont {A.}~\bibnamefont {van
  Elst}}, \ and\ \bibinfo {author} {\bibfnamefont {H.}~\bibnamefont
  {Schulz-Baldes}},\ }\href {\doibase 10.1063/1.530758} {\bibfield  {journal}
  {\bibinfo  {journal} {Journal of Mathematical Physics}\ }\textbf {\bibinfo
  {volume} {35}},\ \bibinfo {pages} {5373} (\bibinfo {year}
  {1994})}\BibitemShut {NoStop}%
\bibitem [{\citenamefont {Kitaev}(2006)}]{Kitaev_2006}%
  \BibitemOpen
  \bibfield  {author} {\bibinfo {author} {\bibfnamefont {A.}~\bibnamefont
  {Kitaev}},\ }\href
  {https://www.sciencedirect.com/science/article/pii/S0003491605002381?via%3Dihub}
  {\bibfield  {journal} {\bibinfo  {journal} {Annals of Physics}\ }\textbf
  {\bibinfo {volume} {321}},\ \bibinfo {pages} {2 } (\bibinfo {year} {2006})},\
  \bibinfo {note} {january Special Issue}\BibitemShut {NoStop}%
\bibitem [{\citenamefont {Prodan}(2009)}]{Prodan_2009}%
  \BibitemOpen
  \bibfield  {author} {\bibinfo {author} {\bibfnamefont {E.}~\bibnamefont
  {Prodan}},\ }\href {\doibase 10.1103/PhysRevB.80.125327} {\bibfield
  {journal} {\bibinfo  {journal} {Phys. Rev. B}\ }\textbf {\bibinfo {volume}
  {80}},\ \bibinfo {pages} {125327} (\bibinfo {year} {2009})}\BibitemShut
  {NoStop}%
\bibitem [{\citenamefont {Prodan}\ \emph {et~al.}(2010)\citenamefont {Prodan},
  \citenamefont {Hughes},\ and\ \citenamefont
  {Bernevig}}]{Prodan_Hughes_Bernevig_2010}%
  \BibitemOpen
  \bibfield  {author} {\bibinfo {author} {\bibfnamefont {E.}~\bibnamefont
  {Prodan}}, \bibinfo {author} {\bibfnamefont {T.~L.}\ \bibnamefont {Hughes}},
  \ and\ \bibinfo {author} {\bibfnamefont {B.~A.}\ \bibnamefont {Bernevig}},\
  }\href {\doibase 10.1103/PhysRevLett.105.115501} {\bibfield  {journal}
  {\bibinfo  {journal} {Phys. Rev. Lett.}\ }\textbf {\bibinfo {volume} {105}},\
  \bibinfo {pages} {115501} (\bibinfo {year} {2010})}\BibitemShut {NoStop}%
\bibitem [{\citenamefont {Prodan}(2011)}]{Prodan_review_2011}%
  \BibitemOpen
  \bibfield  {author} {\bibinfo {author} {\bibfnamefont {E.}~\bibnamefont
  {Prodan}},\ }\href {http://stacks.iop.org/1751-8121/44/i=11/a=113001}
  {\bibfield  {journal} {\bibinfo  {journal} {Journal of Physics A:
  Mathematical and Theoretical}\ }\textbf {\bibinfo {volume} {44}},\ \bibinfo
  {pages} {113001} (\bibinfo {year} {2011})}\BibitemShut {NoStop}%
\bibitem [{\citenamefont {Drabkin}\ \emph {et~al.}()\citenamefont {Drabkin},
  \citenamefont {De~Nittis},\ and\ \citenamefont
  {Schulz-Baldes}}]{Schulz-Baldes_BdG}%
  \BibitemOpen
  \bibfield  {author} {\bibinfo {author} {\bibfnamefont {M.}~\bibnamefont
  {Drabkin}}, \bibinfo {author} {\bibfnamefont {G.}~\bibnamefont {De~Nittis}},
  \ and\ \bibinfo {author} {\bibfnamefont {H.}~\bibnamefont {Schulz-Baldes}},\
  }\href {https://arxiv.org/abs/1310.0207} {\ }\Eprint
  {http://arxiv.org/abs/1310.0207} {arXiv:1310.0207 [math-ph]} \BibitemShut
  {NoStop}%
\bibitem [{\citenamefont {Elgart}\ \emph {et~al.}(2005)\citenamefont {Elgart},
  \citenamefont {Graf},\ and\ \citenamefont {Schenker}}]{Graf_2005}%
  \BibitemOpen
  \bibfield  {author} {\bibinfo {author} {\bibfnamefont {A.}~\bibnamefont
  {Elgart}}, \bibinfo {author} {\bibfnamefont {G.}~\bibnamefont {Graf}}, \ and\
  \bibinfo {author} {\bibfnamefont {J.}~\bibnamefont {Schenker}},\ }\href
  {\doibase 10.1007/s00220-005-1369-7} {\bibfield  {journal} {\bibinfo
  {journal} {Communications in Mathematical Physics}\ }\textbf {\bibinfo
  {volume} {259}},\ \bibinfo {pages} {185} (\bibinfo {year}
  {2005})}\BibitemShut {NoStop}%
\bibitem [{\citenamefont {Avron}\ \emph {et~al.}(1994)\citenamefont {Avron},
  \citenamefont {Seiler},\ and\ \citenamefont
  {Simon}}]{Avron_Seiler_Simon_1994}%
  \BibitemOpen
  \bibfield  {author} {\bibinfo {author} {\bibfnamefont {J.}~\bibnamefont
  {Avron}}, \bibinfo {author} {\bibfnamefont {R.}~\bibnamefont {Seiler}}, \
  and\ \bibinfo {author} {\bibfnamefont {B.}~\bibnamefont {Simon}},\ }\href
  {\doibase 10.1007/BF02102644} {\bibfield  {journal} {\bibinfo  {journal}
  {Commun. Math. Phys.}\ }\textbf {\bibinfo {volume} {159}},\ \bibinfo {pages}
  {399} (\bibinfo {year} {1994})}\BibitemShut {NoStop}%
\bibitem [{\citenamefont {Simon}(2015)}]{Simon_Operator_Theory}%
  \BibitemOpen
  \bibfield  {author} {\bibinfo {author} {\bibfnamefont {B.}~\bibnamefont
  {Simon}},\ }\href@noop {} {\emph {\bibinfo {title} {Operator Theory: A
  Comprehensive Course in Analysis, Part 4}}}\ (\bibinfo  {publisher} {AMS},\
  \bibinfo {year} {2015})\BibitemShut {NoStop}%
\bibitem [{\citenamefont {Reed}\ and\ \citenamefont
  {Simon}(1975)}]{Reed_Simon_2}%
  \BibitemOpen
  \bibfield  {author} {\bibinfo {author} {\bibfnamefont {M.}~\bibnamefont
  {Reed}}\ and\ \bibinfo {author} {\bibfnamefont {B.}~\bibnamefont {Simon}},\
  }\href@noop {} {\emph {\bibinfo {title} {Methods of Modern Mathematical
  Physics, Vol. 2}}}\ (\bibinfo  {publisher} {Academic Press},\ \bibinfo {year}
  {1975})\BibitemShut {NoStop}%
\bibitem [{\citenamefont {Galindo}\ and\ \citenamefont
  {Pascual}(1991)}]{Galindo_Pascual_2}%
  \BibitemOpen
  \bibfield  {author} {\bibinfo {author} {\bibfnamefont {A.}~\bibnamefont
  {Galindo}}\ and\ \bibinfo {author} {\bibfnamefont {R.}~\bibnamefont
  {Pascual}},\ }\href@noop {} {\emph {\bibinfo {title} {Quantum Mechanics
  II}}}\ (\bibinfo  {publisher} {Springer},\ \bibinfo {year}
  {1991})\BibitemShut {NoStop}%
\bibitem [{\citenamefont {Lieb}\ and\ \citenamefont
  {Robinson}(1972)}]{Lieb_Robinson:1972}%
  \BibitemOpen
  \bibfield  {author} {\bibinfo {author} {\bibfnamefont {E.~H.}\ \bibnamefont
  {Lieb}}\ and\ \bibinfo {author} {\bibfnamefont {D.~W.}\ \bibnamefont
  {Robinson}},\ }\href {\doibase 10.1007/BF01645779} {\bibfield  {journal}
  {\bibinfo  {journal} {Commun. math. Phys.}\ }\textbf {\bibinfo {volume}
  {28}},\ \bibinfo {pages} {251} (\bibinfo {year} {1972})}\BibitemShut
  {NoStop}%
\bibitem [{\citenamefont {Hastings}()}]{Hastings_Les_Houches:2010}%
  \BibitemOpen
  \bibfield  {author} {\bibinfo {author} {\bibfnamefont {M.~B.}\ \bibnamefont
  {Hastings}},\ }\href {https://arxiv.org/abs/1008.5137} {\ }\Eprint
  {http://arxiv.org/abs/1008.5137} {arXiv:1008.5137 [math-ph]} \BibitemShut
  {NoStop}%
\bibitem [{\citenamefont {Russomanno}\ \emph {et~al.}(2012)\citenamefont
  {Russomanno}, \citenamefont {Silva},\ and\ \citenamefont
  {Santoro}}]{Russomanno_Silva_2012}%
  \BibitemOpen
  \bibfield  {author} {\bibinfo {author} {\bibfnamefont {A.}~\bibnamefont
  {Russomanno}}, \bibinfo {author} {\bibfnamefont {A.}~\bibnamefont {Silva}}, \
  and\ \bibinfo {author} {\bibfnamefont {G.~E.}\ \bibnamefont {Santoro}},\
  }\href {\doibase 10.1103/PhysRevLett.109.257201} {\bibfield  {journal}
  {\bibinfo  {journal} {Phys. Rev. Lett.}\ }\textbf {\bibinfo {volume} {109}},\
  \bibinfo {pages} {257201} (\bibinfo {year} {2012})}\BibitemShut {NoStop}%
\bibitem [{\citenamefont {Abanin}\ \emph {et~al.}(2015)\citenamefont {Abanin},
  \citenamefont {De~Roeck},\ and\ \citenamefont {Huveneers}}]{Abanin_2015}%
  \BibitemOpen
  \bibfield  {author} {\bibinfo {author} {\bibfnamefont {D.~A.}\ \bibnamefont
  {Abanin}}, \bibinfo {author} {\bibfnamefont {W.}~\bibnamefont {De~Roeck}}, \
  and\ \bibinfo {author} {\bibfnamefont {F.~m.~c.}\ \bibnamefont {Huveneers}},\
  }\href {\doibase 10.1103/PhysRevLett.115.256803} {\bibfield  {journal}
  {\bibinfo  {journal} {Phys. Rev. Lett.}\ }\textbf {\bibinfo {volume} {115}},\
  \bibinfo {pages} {256803} (\bibinfo {year} {2015})}\BibitemShut {NoStop}%
\bibitem [{\citenamefont {Abanin}\ \emph {et~al.}(2017)\citenamefont {Abanin},
  \citenamefont {De~Roeck}, \citenamefont {Ho},\ and\ \citenamefont
  {Huveneers}}]{Abanin_2017}%
  \BibitemOpen
  \bibfield  {author} {\bibinfo {author} {\bibfnamefont {D.}~\bibnamefont
  {Abanin}}, \bibinfo {author} {\bibfnamefont {W.}~\bibnamefont {De~Roeck}},
  \bibinfo {author} {\bibfnamefont {W.~W.}\ \bibnamefont {Ho}}, \ and\ \bibinfo
  {author} {\bibfnamefont {F.}~\bibnamefont {Huveneers}},\ }\href {\doibase
  10.1007/s00220-017-2930-x} {\bibfield  {journal} {\bibinfo  {journal}
  {Communications in Mathematical Physics}\ }\textbf {\bibinfo {volume}
  {354}},\ \bibinfo {pages} {809} (\bibinfo {year} {2017})}\BibitemShut
  {NoStop}%
\bibitem [{\citenamefont {Titum}\ \emph {et~al.}(2015)\citenamefont {Titum},
  \citenamefont {Lindner}, \citenamefont {Rechtsman},\ and\ \citenamefont
  {Refael}}]{Refael:2015}%
  \BibitemOpen
  \bibfield  {author} {\bibinfo {author} {\bibfnamefont {P.}~\bibnamefont
  {Titum}}, \bibinfo {author} {\bibfnamefont {N.~H.}\ \bibnamefont {Lindner}},
  \bibinfo {author} {\bibfnamefont {M.~C.}\ \bibnamefont {Rechtsman}}, \ and\
  \bibinfo {author} {\bibfnamefont {G.}~\bibnamefont {Refael}},\ }\href
  {\doibase 10.1103/PhysRevLett.114.056801} {\bibfield  {journal} {\bibinfo
  {journal} {Phys. Rev. Lett.}\ }\textbf {\bibinfo {volume} {114}},\ \bibinfo
  {pages} {056801} (\bibinfo {year} {2015})}\BibitemShut {NoStop}%
\bibitem [{\citenamefont {Ho}\ and\ \citenamefont {Abanin}()}]{Abanin_2016}%
  \BibitemOpen
  \bibfield  {author} {\bibinfo {author} {\bibfnamefont {W.~W.}\ \bibnamefont
  {Ho}}\ and\ \bibinfo {author} {\bibfnamefont {D.~A.}\ \bibnamefont
  {Abanin}},\ }\href {https://arxiv.org/abs/1611.05024} {\ }\Eprint
  {http://arxiv.org/abs/1611.05024} {arXiv:1611.05024 [cond-mat]} \BibitemShut
  {NoStop}%
\bibitem [{\citenamefont {Sch\"uler}\ and\ \citenamefont
  {Werner}(2017)}]{Werner:2017}%
  \BibitemOpen
  \bibfield  {author} {\bibinfo {author} {\bibfnamefont {M.}~\bibnamefont
  {Sch\"uler}}\ and\ \bibinfo {author} {\bibfnamefont {P.}~\bibnamefont
  {Werner}},\ }\href {\doibase 10.1103/PhysRevB.96.155122} {\bibfield
  {journal} {\bibinfo  {journal} {Phys. Rev. B}\ }\textbf {\bibinfo {volume}
  {96}},\ \bibinfo {pages} {155122} (\bibinfo {year} {2017})}\BibitemShut
  {NoStop}%
\end{thebibliography}%

\end{document}